# Bridging thermal and electrical transport in dielectric nanostructure based polar colloids

Purbarun Dhar [a, ▲], Soujit Sen Gupta [b], Arvind Pattamatta [a,*] and Sarit K. Das [a,$]

[a] Department of Mechanical Engineering, Indian Institute of Technology, Madras,

Chennai – 600 036

[b] Department of Chemistry, Indian Institute of Technology, Madras,

Chennai – 600 036

[▲]E–mail: pdhar1990@gmail.com

[*]E–mail: arvindp@iitm.ac.in

[$]Corresponding author: skdas@iitm.ac.in

Telephone: +91-44-2257 4655

Fax: +91-44-2257 8040

# Abstract

Heat and charge transport characteristics of nanocolloids have been bridged from fundamental analysis. The relationship between the two transport phenomena in dielectric nanostructure based polar colloids has been quantitatively presented. An extensional intuitive analogy to the Wiedemann–Franz law has been drawn. Derived from the fact that mobile electrons transport both heat and charge within metallic crystal structure, the analogy can be extended to nanocolloids, wherein the dispersed population act as the major transporter. The analogy allows modeling the relationship between the two phenomena and sheds more insight and conclusive evidence that nanoparticle traversal within the fluid domain is the main source of augmented transport phenomena exhibit by nanocolloids. Important factors such as the thermal and dielectric responses of the nanocolloid can be quantified and bridged through the semi–analytical formalism. The theoretical analysis has been validated against experimental data and variant scientific literature and good accuracy has been observed.

Thermal and electrical transport in nanocolloids[1, 2] have been intensively studied over the past decade. The phenomena have been experimentally observed[3, 4, 5, 6, 7, 8] and mathematically modeled[9, 10, 11, 12, 13]. Although variant mechanisms have been proposed to explain the transport behavior[14, 15, 16], no light on the existence of possible coupled nature of the two transport phenomena has been shed. Thermal and charge transport in nanocolloids can be hypothesized to be borne of a common mechanism: the streaming of nanoparticles; either due to thermal fluctuations (Brownian mobility, thermophoresis, etc.) or electrophoresis. The fundamental pillar for this hypothesis rests on the fact that nanocolloids mimic metals, where the conduction band electrons act as transporters of energy across the crystal structure. Similarly, the particles act as mobile transporters across the fluidic medium. The approach in the present formulation is based on extending the Wiedemann-Franz law[17] to nanocolloids. Mobile electrons transfer thermal energy via internal storage and charge due to intrinsic property. Likewise, nanoparticles acquire heat by interactions with fluid molecules and neighboring particles and charge is generated due to formation of the Electric Double Layer (EDL) at the particle–fluid interface. The two systems are in essence similar, and can be hypothesized to obey similar physical models. The Wiedemann-Franz law is expressible as[17]

$$\frac{k}{\sigma T} = L \qquad (1)$$

'$k$', '$\sigma$', '$T$' and '$L$' represent the thermal conductivity, electrical conductivity, absolute temperature and the Lorentz constant. However, this basic formulation fails to bridge the two transport parameters since fluidic systems are more complex in structural and aggregation aspects. Unlike a metallic lattice, wherein the atomic positions are fixed in spatio–temporal coordinates, a fluidic system is mobile. Although a matrix analogy can be conceived, the fluid molecules move about their mean positions with spatio–temporally variant velocities. Therefore Eq. 1 cannot be applied without specific modifications. A pictographic representation of the conceived matrix hypothesis is illustrated in Fig. (1). Furthermore, colloids are phase mixtures and exhibit physical properties dependent on the morphological and physical characteristics of the dispersed, making the mathematical representation further complex. In the present article, a simplistic route to determine the bridging parameter of the two transport mechanisms has been proposed. An intuitive approach, semi-empirical by nature, has been developed to extend Eqn. (1) to nanocolloids. Since nanoparticles are not fundamental entities like electrons, the equation requires modifications so as to implement their thermal and dielectric properties. Furthermore, the concentration and viscous resistance experienced by the mobile particles require explicit incorporation within the formalism.

## FIGURE 1

The approach involves intuitively choosing the right transport parameters so as to normalize Eqn. (1) such that both terms on either sides of the equivalence operator convert to pure functions of transport properties of the colloidal system constituents, thereby bridging its thermal and electrical transport properties. In this process, the diametric magnitude, thermal conductivity, relative permittivity and concentration of the dispersed phase and the dynamic viscosity of the fluid are utilized. Normalization is performed based on the relation of the concerned property to the thermal and electrical conductivities of the nanocolloid. The particle diameter '$d_{np}$' maintains an inverse relationship with the thermal conductivity of the nanocolloid[1, 4, 8, 11, 16], '$k_c$'. However, the thermal conductivity of the nanocolloid holds a direct relationship with the thermal conductivity of the particulate phase[7, 18, 19] '$k_{np}$' and the viscosity of the fluidic phase[9, 11, 19, 20], '$\mu$'. Similarly, the electrical conductivity of the colloid, '$\sigma_c$', exhibits direct dependence on the relative permittivity of the dispersed phase[5, 9, 21, 22], '$\varepsilon_{r,np}$', and an inverse relation to the viscosity of the fluidic phase[9, 21]. However, the dependencies of thermal conductivity and electrical conductivity on the viscosity of the fluidic phase are non–equivalent[9, 18] and an exponent is required to map the exact response. The viscous resistance to the mobility of the particulate phase is of utmost importance since the transport of energy is mediated by diffusivity of particles, which is controlled by the viscous drag[11]. The final normalized form, in terms of the colloid concentration, '$\varphi$', is expressed as

$$\left(\frac{k_c}{\sigma_c T}\right)\left(\frac{d_{np}\varepsilon_{r,np}}{k_{np}\mu_f^m}\right)\varphi = C \qquad (2)$$

'$C$' is a variable unlike the Lorenz constant[23] in the Wiedemann–Franz law and can be deduced from the physical properties of the colloid. Detailed analysis of experimental data[8, 19] suggests that essentially all colloids follow a general trend. Based on this, in simplistic form, it can be expressed as a function of concentration as

$$C = d_{\text{inf}}\varphi^{\tau_r} \qquad (3)$$

where, '$d_{inf}$' and '$\tau_r$' are proposed as the diameter of influence for the particle-fluid system and the thermal response index of the particle with respect to the fluid respectively. From the proposed expression for '$C$', it can be seen that the heat and charge transport parameters of nanocolloids is governed by the concentration of the dispersed phase. The dependence is further governed at the particle–fluid interface[15] by thermal and dielectric properties of the particulate and fluid phases and their interactions, which can be mathematically explained in form of the variables '$d_{inf}$' and '$\tau_r$'. In a stable colloid individual particles are shrouded by an EDL[24] whose diametric magnitude depends upon factors such as the surface charge density of the particles, the dielectric properties of the phases, concentration[9], etc. Intrinsically, the mobility of the dispersed phase is governed by the extent to which the EDL extends from the surface of the particle[24]. Also, due to the combined effects of thermal and electrodiffusion in presence of external ield, the particle exists in a constant state of randomness that can be described on the basis of Langevin dynamics[25]. Thereby, it is difficult to conceive the particle at a fixed position in space and suitable metrics must be defined to ascertain the probabilistic position of the particle. This is important since the positional aspect gives a clear picture of the diffusive randomness, and thus the transport capabilities of the system. In the present context, it is a necessity to hypothesize '$d_{inf}$' for the particle. The '$d_{inf}$' may be defined as the effective diameter of an imaginary sphere in the fluidic phase wherein the effects of interplay of van der Waals forces, electrostatic repulsions and thermo diffusion between the particle and neighboring fluid molecules is confined. It is such that the forces exerted by the nanoparticle on fluid molecules at a distance of '$d_{inf}$' or more from the center of the nanoparticle under observation are negligibly minimal, i.e., all electro–thermal fluctuations of the particle are confined within the '$d_{inf}$'. It is calculated based on the criterion of stability of the suspended particle in conjunction with diffusion within the fluid. Formulated based on numerous experimental observations, '$d_{inf}$' is mathematically expressed as

$$d_{\inf} = d_{np}\left(1 + \frac{2\delta}{\pi}\right) = d_{np}\left(1 + \xi\right) \qquad (4)$$

'$\xi$' denotes the enhancement factor by which the electro–thermal influence of the particle is perceived by the polar fluid when stabilized in it. It is difficult to model the variant '$\delta$' via direct mathematical analysis and has been determined through intuitive analysis of a voluminous body of experimental data. '$\delta$' has been proposed as the thermo–dielectric stability correlation factor and provides a perspective on the stability of the particulate phase under the influence of thermal diffusion in presence of the EDL. '$\delta$' has been observed to have a general mathematical structure for all nanomaterials in variant polar fluids, expressible as

$$\delta = \frac{k_{np}^{\chi} \varepsilon_{r,np}^{\psi}}{\rho_{np}} \qquad (5)$$

'$\chi$' and '$\psi$' represent integral exponents which govern the thermal and electrical diffusive components of the dispersed phase within the fluid. The magnitudes of the two variables provide a picture of the thermal and electrical response of the particles. The density of the particulate phase, '$\rho_{np}$', is also an important governing parameter since it decides the stability of the dispersed phase within the fluidic phase, as derivable from the Stokes–Einstein's formulation. As intuition predicts, a denser dispersed phase will lead to a less stable system. Also, higher values of thermal conductivity and dielectric constant will lead to higher diffusive response of the particle within the fluid, leading to higher '$d_{inf}$', and therefore the form of Eqn. (5) is physically valid.

'$\tau_r$' is a governing parameter dependent on the thermal coupling between the dispersed and the fluidic phase. It is diffusion dominated and quantifies the thermal state of the colloidal system and provides insight into the randomness of the particulate phase. It can be expressed mathematically as

$$\tau_r = \frac{k_{np}}{k_f Tn} \qquad (6)$$

It is a function of the ratio of thermal conductivity of the particulate phase to that of the fluidic phase, '$k_f$'. This has a similitude to the classical formalisms based on the Enskog equation with simplified scattering parameter[25]. The analytical solution for thermal conductivity of a low viscosity liquid from the simplified Enskog equation contains a similar term where the ratio of the thermal conductivity of the fluidic phase of interest to that of a dilute gas phase is important. In essence the ratio of transport properties of a semi–condensed phase to a gas phase system becomes important. Similarly, here, the ratio of transport parameters of a condensed phase system to a semi–condensed phase is important to quantify transport within the colloidal system. Higher the particle thermal conductivity with respect to the fluidic phase, higher is the response of the colloid to temperature. Higher the ratio, the ability of the dispersed phase to extract more heat from a given set of spatial coordinates at a given temporal instant is higher than its fluidic counterparts. Due to larger extraction of thermal energy and distribution to other spatial locations owing to subsequent diffusion, the thermal conductivity of such colloids are better augmented than for colloids with lower ratios of $k_{np}/k_f$. This is in fact established by detailed experimental reports[4, 26]. '$\tau_r$' being a response quantifier is also a function of a diffusive term and observed to be inversely dependent on the absolute temperature. Interestingly enough, the inverse of the absolute temperature is often the qualitative projection of the thermophoretic or the Ludwig–Soret diffusion coefficient of a liquid based dispersion system[27]. This sheds light onto the physical aspects of thermal and electrical transport involved in nanocolloids. The presence of the thermodiffusion term signifies that transport is governed by the thermal response of the particulate phase. The transport parameters, being coupled as explained through the preceding analysis, leads to charge conduction. This provides a quantitative reasoning of the similar response of the thermal and electrical transport properties of nanocolloids to temperature[3, 9].

The viscous component within the formulation (Eqn. (2)) is index corrected to compensate for the difference in viscous characteristics of the nanocolloid from that of the fluid.. It comprises of raising the dynamic viscosity of the fluidic phase to an index, '*m*', proposed as a viscosity correction index. It is noteworthy that the present approach is valid for fluidic systems where the thermal conductivity and viscosity can be predicted fairly accurately from the analytical solutions of the scattering simplified Enskog equation[25]. The solution for the dynamic viscosity of a low viscosity liquid from the simplified Enskog equation contains a similar term where the ratio of the viscosity of the concerned fluid to that of a dilute gaseous phase is important. Corrective approach on the viscous behavior of the dilute gas can provide fairly accurate predictions of the dynamic viscosity of liquid phases with low or moderate viscosities. The analytical solution of Enskog equation therefore cannot determine the magnitude of viscosity of highly viscous fluids since the intermolecular attractive potentials are much higher compared to a dilute gas. Therefore, the present formulation cannot bridge the transport phenomena in highly viscous fluids such as glycerol, viscous mineral oils, etc. however, it is interesting to note that for fluids with viscosities moderately higher (within an order of magnitude) than systems such as water or similar fluids, the transport parameters can be bridged by further correcting the viscosity component utilizing the proposed methodology.

The main philosophy behind the approach lies in correcting the viscous characteristics of the fluid so as to accommodate the fluid within the limits of low viscosity so that intermolecular potentials do not hinder the transport processes. It involves correcting the magnitude of the dynamic viscosity effectively by the use of the parameter '$m$'. It qualitatively maps the inter–molecular randomness and interactivities of the fluid medium under observation with respect to that of a fluid medium whose viscosity can be predicted from the modified Enskog equation. It can be expressed as

$$m = 1 - \frac{f\varphi}{n} \qquad (7)$$

As observable from Eqn. (3), '$m$' is a function of the concentration of the particulate phase, and therefore, the corrective approach is only possible for colloids and not for the base fluids. This is representative of the fact that it maps the motion of the dispersed phases caused by the thermal motion of the fluid molecules and then corrects the viscous forces within the colloid. The factor '$n$' correlates the molecular randomness of a fluidic media to that of another fluid. In case the viscosity of the reference fluid can be derived based on Enskog's formalism, the above approach can accurately correct the viscous properties of the subject fluid. '*n*' is expressible as

$$n = \frac{E - E_{m,f}}{E - E_{m,ref}} = \frac{T - T_{m,f}}{T - T_{m,ref}} \qquad (8)$$

'*n*' correlates the thermal energy ($E = k_B T$, $k_B$ is the Boltzmann constant) of the fluidic system with respect to that of another fluidic system which closely follows the predictions of the Enskog equation (subscripted as '*ref*'). Water has been utilized as such a reference fluid. If the fluid of interest is also water, the magnitude of '*n*' is simply unity. For other fluidic phases, where the thermal states are not clearly understood due to their divergence from the Enskog equation, the factor can be utilized to scale the viscous forces accurately. From thermodynamic point of view, the absolute temperature of a system can provide a qualitative picture of the thermal randomness of the system. At the solid–liquid first order phase transition point, the fluidic system is converted to solid phase and the mobility of the fluidic molecules ceases to exist. It is at this state that the fluidic system with respect to the particulate phase can be assigned a reference state of zero molecular mobility. Thereby, the difference of the temperature of interest with respect to the melting point temperature provides an assessment of the thermal state of the system. The ratio of the difference to that of the difference for the reference fluid thereby provides an index to correlate the thermal states of the two fluids, which is used to scale the viscous forces at a given temperature.

However, the expression in Eqn. (3) also requires a mathematical expression for the normalizing variable '*f*'. Viscosity scaling requires a basic index that can provide a quantitative magnitude of the degree to which the fluidic phase is more viscous than the reference fluid. Essentially a viscosity rank correlation, it has been denoted by '*β*'. Since viscosity is a strong function of temperature, the magnitude by which a fluid is more viscous than another changes non–linearly with temperature. Thereby, a rank correlative approach to determine '$\beta$' is assorted to. The mathematical expression for '$\beta$' is expressed as

$$\beta = \left[ \frac{\prod_{T=T_1}^{T=T_2} \mu_{T,f}}{\prod_{T=T_1}^{T=T_2} \mu_{T,ref}} \right] \quad (9)$$

The approach involves determining the product of the viscosities of the fluidic phase at particular temperatures within a pre–determined temperature range, at a fixed sampling interval (indicated by the numerator of Eqn. (5)). The same procedure is followed for the reference fluid, whose viscosity can be predicted within appreciable limits by the analytical Enskog approach (indicated by the denominator of Eqn. (5)). The point of interest lies in the data sampling frequency involved since the accuracy of predictability of the viscous forces depends on the value of '$\beta$'. The normalizing factor '$f$' is derived based on exponent based reduction of '$\beta$'. Mathematically, the expression in Eqn. (6) is found to be the most consistent approach, as observed from data analysis of detailed experimental results.

$$f = \beta^{1/2\theta} \quad (10)$$

The physical significance of the normalizing factor in essence is simple. '$\beta$', obtained from a rank correlative approach provides the degree to which the fluidic system is more viscous than the reference system over a preset range of temperatures. However, '$\beta$' requires normalization and needs to be expressed for a specific mean temperature within the utilized temperature range. Thereby, an equivalent geometric mean is utilized to obtain the factor '$f$'. The parameter involved in the geometric normalization process requires the effective temperature spectral range utilized in the ranking process and the temperature sampling magnitude utilized. The mathematical expression for the same is expressed as

$$\theta = \frac{T_2 - T_1}{\Delta T_{sampling}} \qquad (11)$$

The present formulation has been validated with a large volume of data. Thermal conductivity data from experiments performed and literature have been utilized to predict electrical conductivity data and compared with experimental values. The exact alternate; wherein measured electrical conductivity values have been utilized to predict thermal conductivity and compared with experimental results; has also been presented. Figs. 2, 3, 4, 5 and 6 illustrate the predictability of the model against experimental data.

**FIGURE 2**

**FIGURE 3**

**FIGURE 4**

**FIGURE 5**

**FIGURE 6**

To infer, an analytical, semi–empirical formulation has been proposed based on observational and intuitive analyses of experimental data in order to bridge the thermal and charge transport phenomena in dielectric particle based polar nanocolloids. The formulation allows deducing the thermal transport characteristics based on known charge transport characteristics and vice–versa and has been proposed along the lines of the Wiedemann–Franz law. The approach has been found to predict the transport properties with good accuracy when compared against reported experimental data or against data deduced from reported theoretical models. The formulation sheds insight onto the similar mechanism of heat and charge transport in nanocolloids via diffusive transport of the dispersed phase across the fluidic medium. Further insight onto the roles of Brownian diffusivity, thermophoretic or Ludwig–Soret diffusion and electrophoretic diffusion and their implications on transport characteristics is achieved through the present extensional analogy approach.

# References

1. JA Eastman, SUS Choi, Sheng Li, W Yu, and LJ Thompson, Applied Physics Letters **78** (6), 718 (2001).
2. Stephen US Choi, Journal of Heat Transfer **131** (3), 033106 (2009).
3. Sarit Kumar Das, Nandy Putra, Peter Thiesen, and Wilfried Roetzel, Journal of Heat Transfer **125** (4), 567 (2003).
4. Yimin Xuan and Qiang Li, International Journal of heat and fluid flow **21** (1), 58 (2000).
5. KG Sarojini, Siva V Manoj, Pawan K Singh, T Pradeep, and Sarit K Das, Colloids and Surfaces A: Physicochemical and Engineering Aspects **417**, 39 (2013).
6. M Chiesa and Sarit K Das, Colloids and Surfaces A: Physicochemical and Engineering Aspects **335** (1), 88 (2009).
7. Calvin H. Li and G. P. Peterson, Journal of Applied Physics **99** (8) (2006).
8. Hrishikesh E Patel, Sarit K Das, T Sundararajan, A Sreekumaran Nair, Beena George, and T Pradeep, Applied Physics Letters **83** (14), 2931 (2003).
9. Purbarun Dhar, Arvind Pattamatta, and Sarit K Das, Journal of Nanoparticle Research **16** (10), 1 (2014).
10. Purbarun Dhar, Mohammad Hasan Dad Ansari, Soujit Sen Gupta, V Manoj Siva, T Pradeep, Arvind Pattamatta, and Sarit K Das, Journal of nanoparticle research **15** (12), 1 (2013).
11. S Savithiri, Arvind Pattamatta, and Sarit K Das, Nanoscale research letters **6** (1), 1 (2011).
12. J. B. Hubbard and P. J. Stiles, Journal of Chemical Physics **84** (6955) (1986).
13. Purbarun Dhar, Soujit Sen Gupta, Saikat Chakraborty, Arvind Pattamatta, and Sarit K Das, Applied Physics Letters **102** (16), 163114 (2013).
14. Yuefan Du, Yuzhen Lv, Chengrong Li, Mutian Chen, Yuxiang Zhong, Jianquan Zhou, Xiaoxin Li, and You Zhou, Dielectrics and Electrical Insulation, IEEE Transactions on **19** (3), 770 (2012).
15. W Yu and SUS Choi, Journal of Nanoparticle Research **6** (4), 355 (2004).

16    D Hemanth Kumar, Hrishikesh E Patel, VR Rajeev Kumar, T Sundararajan, T Pradeep, and Sarit K Das, Physical Review Letters **93** (14), 144301 (2004).
17    R Franz and G Wiedemann, Annalen der Physik **165** (8), 497 (1853).
18    Honorine Angue Mintsa, Gilles Roy, Cong Tam Nguyen, and Dominique Doucet, International Journal of Thermal Sciences **48** (2), 363 (2009).
19    Hrishikesh E Patel, T Sundararajan, and Sarit K Das, Journal of Nanoparticle Research **12** (3), 1015 (2010).
20    K. B. Anoop, S. Kabelac, T. Sundararajan, and Sarit K. Das, Journal of Applied Physics **106** (3) (2009).
21    AlinaAdriana Minea and RazvanSilviu Luciu, Microfluid Nanofluid **13** (6), 977 (2012).
22    J. Glory, M. Bonetti, M. Helezen, M. Mayne-L'Hermite, and C. Reynaud, Journal of Applied Physics **103** (9) (2008).
23    Charles Kittel and Paul McEuen, *Introduction to solid state physics*. (Wiley New York, 1976).
24    Brian J Kirby, *Micro-and nanoscale fluid mechanics: transport in microfluidic devices*. (Cambridge University Press, 2010).
25    Gang Chen, *Nanoscale energy transport and conversion: a parallel treatment of electrons, molecules, phonons, and photons*. (Oxford University Press, USA, 2005).
26    R Saidur, KY Leong, and HA Mohammad, Renewable and Sustainable Energy Reviews **15** (3), 1646 (2011).
27    Parola A and Piazza R, Euro. Phys. J. E **15** (255), 263 (2004).
28    S. M. S. Murshed, K. C. Leong, and C. Yang, International Journal of Thermal Sciences **44** (4), 367 (2005).
29    Tae-Keun Hong, Ho-Soon Yang, and C. J. Choi, Journal of Applied Physics **97** (6) (2005).

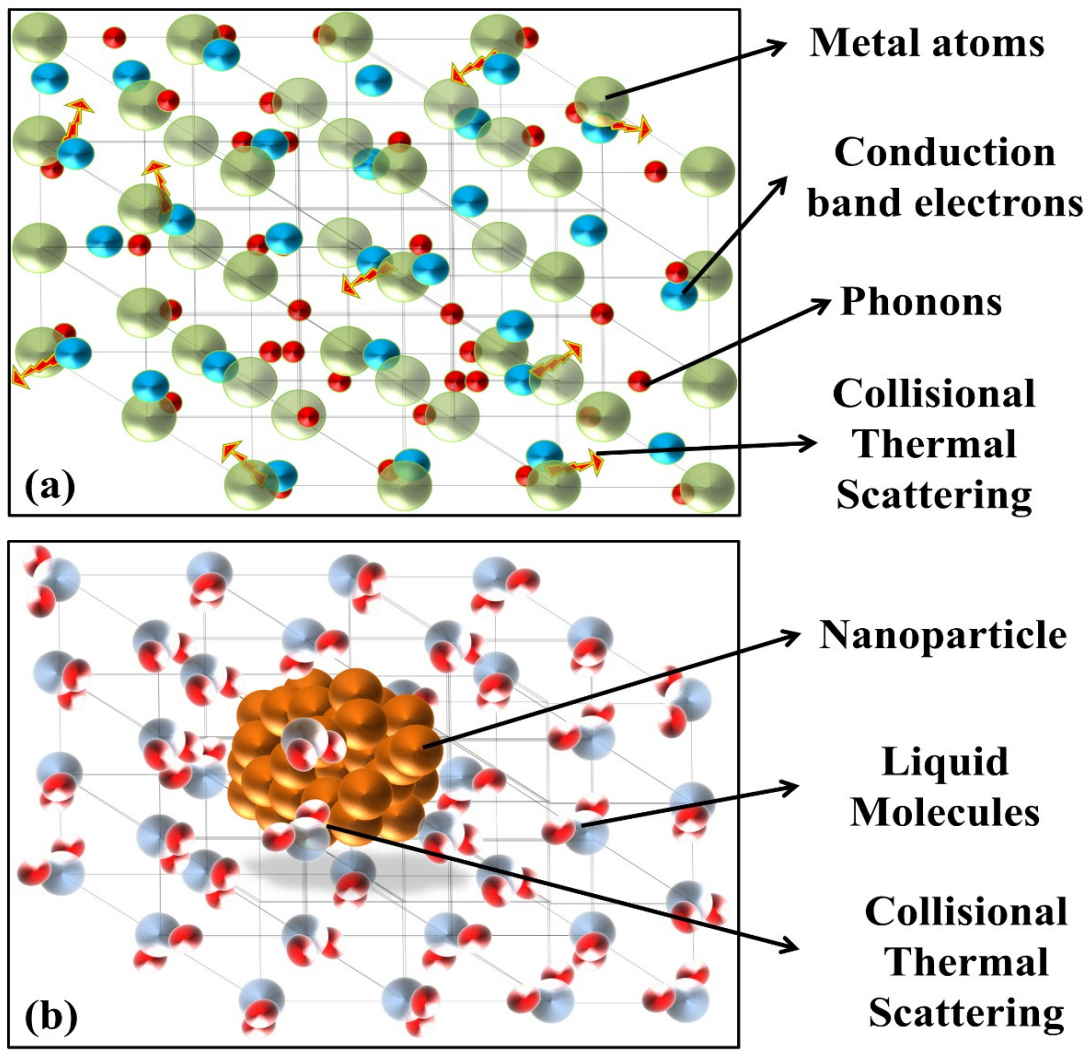

**Figure 1:** Illustration of the extensional analogy drawn from the Wiedemann-Franz law.

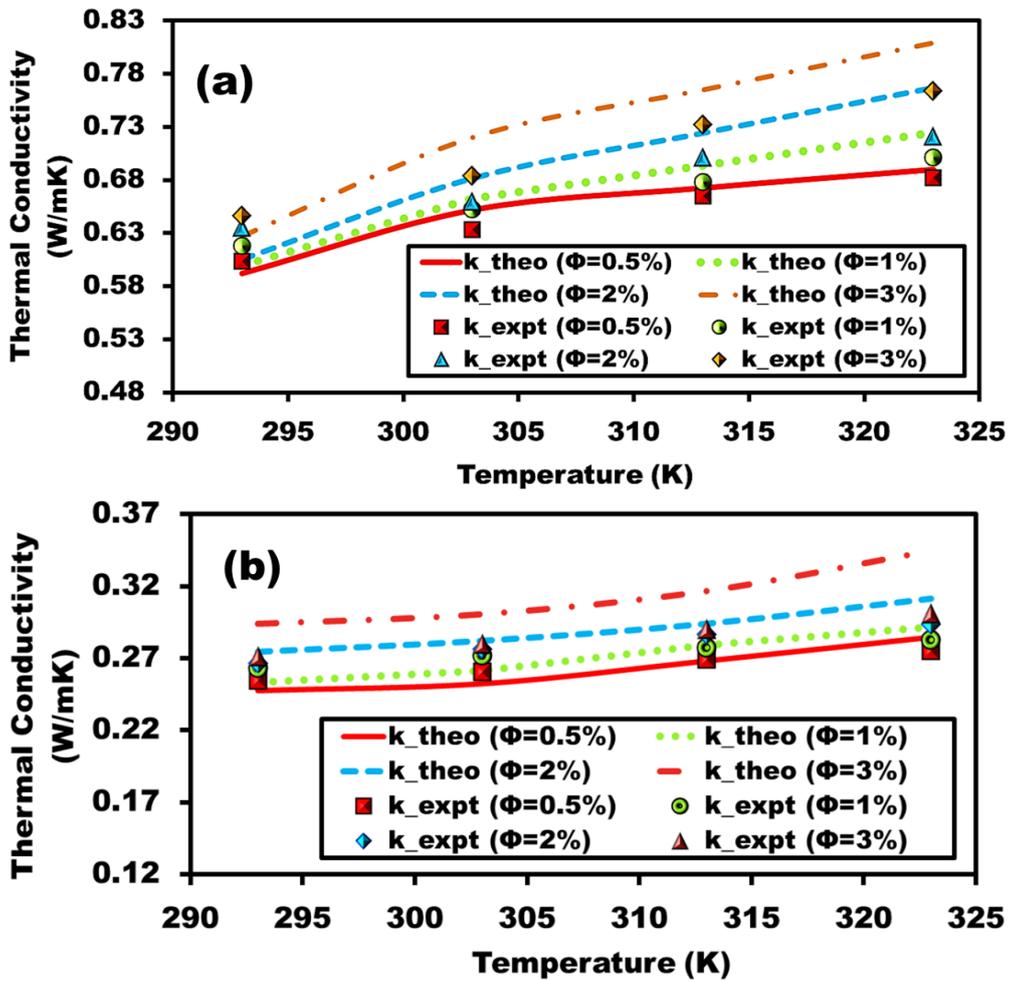

**Figure 2:** Experimental thermal conductivity of nanocolloids[19] (alumina 45 nm in **(a)** water and **(b)** ethylene glycol (EG)) compared against the theoretical values predicted by the present approach. The electrical conductivities of the nanocolloids were experimentally determined and utilized to determine the theoretical values of thermal conductivity in accordance to the formulation.

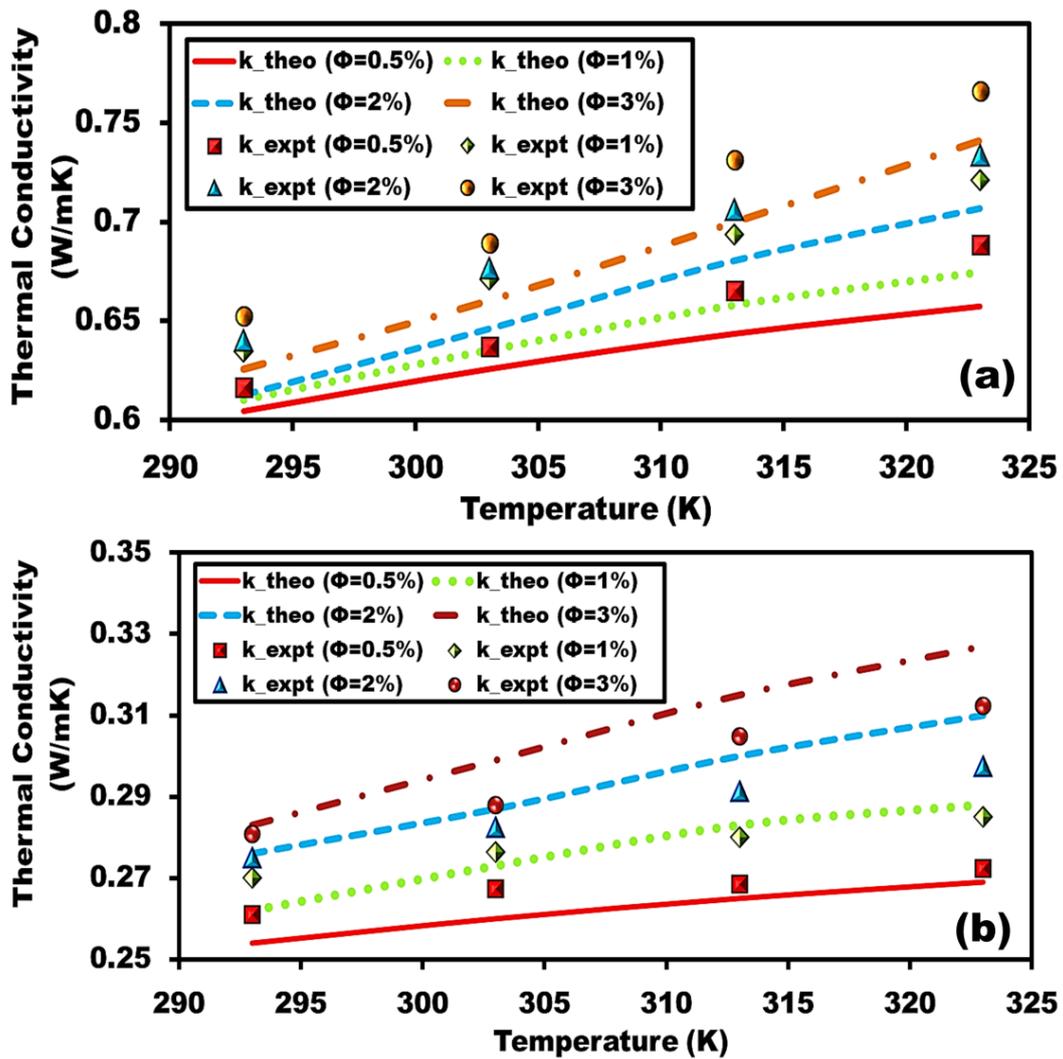

**Figure 3:** Experimental thermal conductivity of nanocolloids[19] (copper oxide 30 nm in **(a)** water and **(b)** EG) compared against the theoretical values predicted by the present approach. The electrical conductivities of the nanocolloids were experimentally determined and utilized to deduce the theoretical values of thermal conductivity in accordance to the formulation.

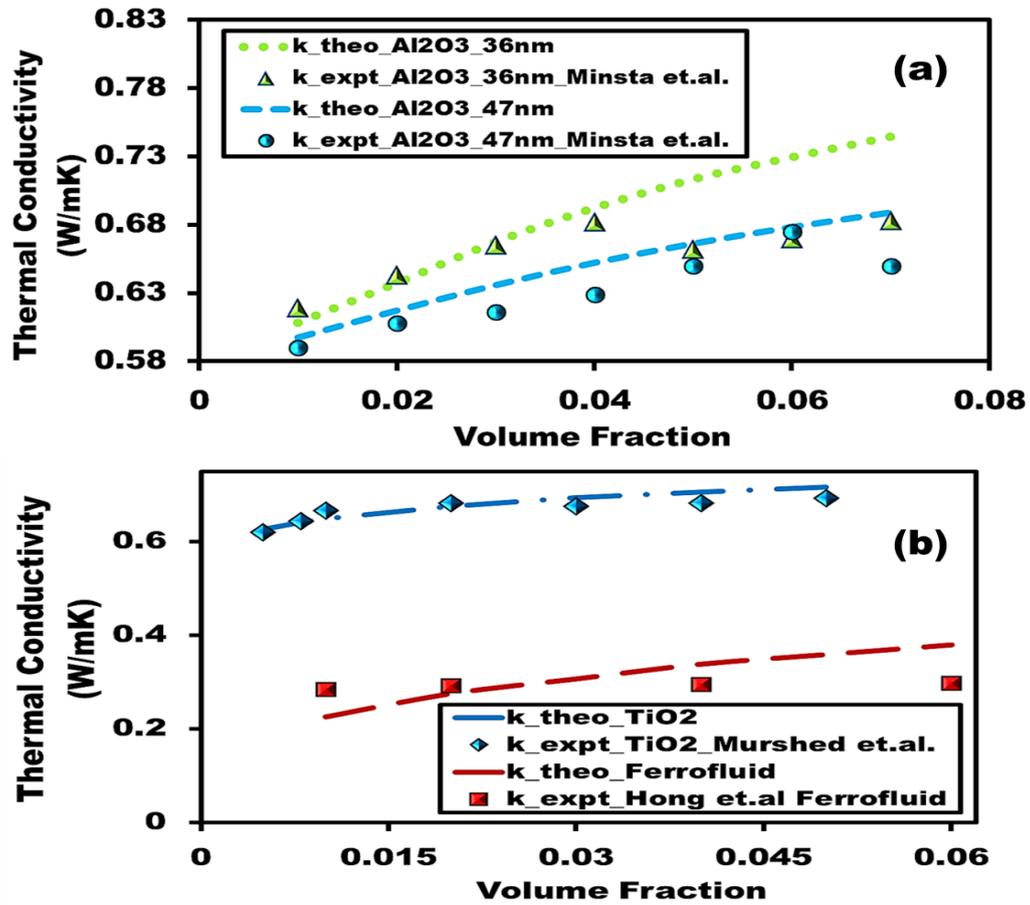

**Figure 4:** Experimental thermal conductivity of variant nanocolloids (from literature) **(a)** Minsta et. al.[18] alumina 36 and 47 nm in water **(b)** Murshed et. al.[28] titania 10 nm in water and Hong et. al.[29] magnetite 10 nm in EG compared with respect to the theoretical values predicted by the present approach. The electrical conductivities of the nanocolloids were theoretically determined[9] and used in the formulation. This is indicative of the general validity of the present model.

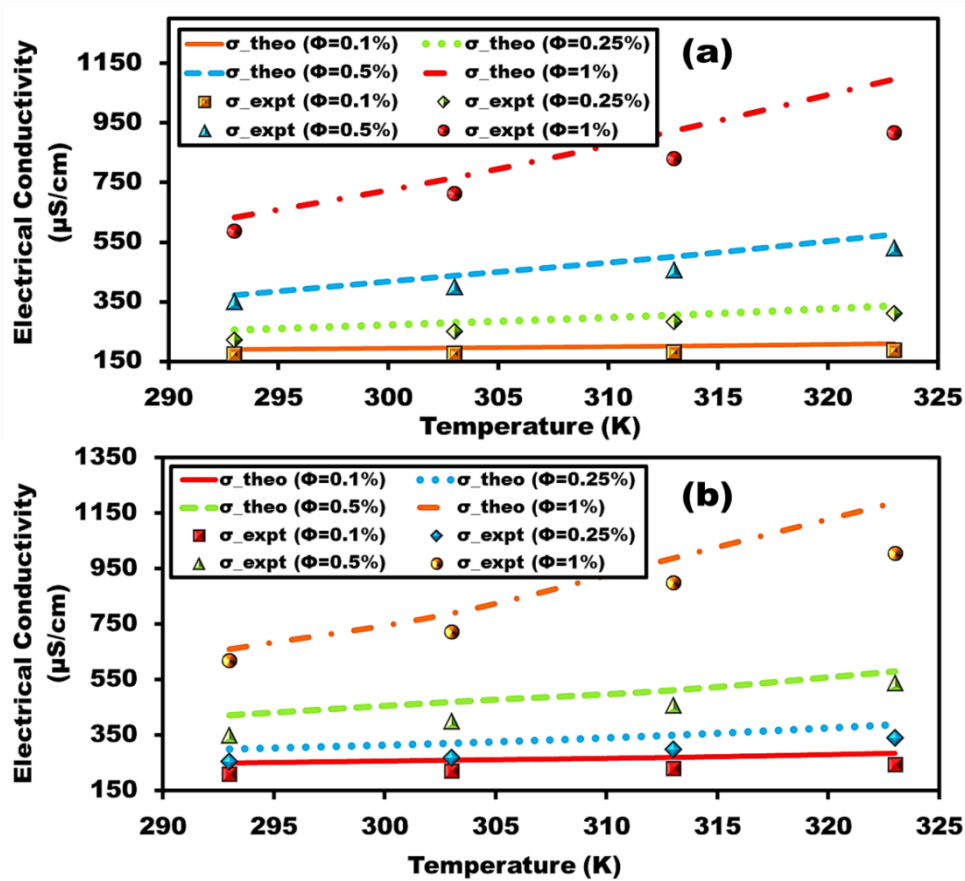

**Figure 5:** Experimental electrical conductivity[5, 9] of variant water based nanocolloids **(a)** alumina 20 nm **(b)** copper oxide 30 nm compared against theoretical values predicted by the present approach. The thermal conductivities of the nanocolloids were determined from experiments.

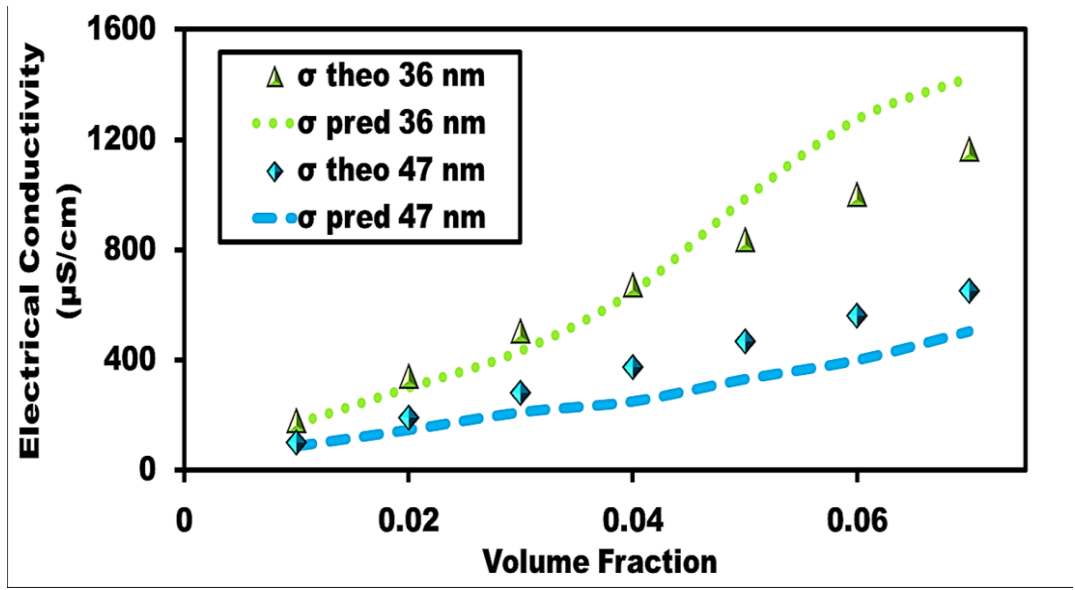

**Figure 6:** Theoretical electrical conductivity[9] of water based alumina nanocolloids compared against values predicted by the present approach. The thermal conductivities of the nanocolloids were adapted from literature[18].